\documentclass[twocolumn,showpacs,prl]{revtex4}
\usepackage{amssymb}
\usepackage{graphicx}
\usepackage{dcolumn}
\usepackage{bm}

\begin{document}

\title{Stabilization of the p-wave superfluid state in an optical lattice}
\author{Y.-J. Han$^{1}$, Y.-H. Chan$^{1}$, W. Yi$^{2}$, A. J. Daley$^{2}$,
S. Diehl$^{2}$, P. Zoller$^{2}$, and L.-M. Duan$^{1}$}
\affiliation{$^{1}$ Department of Physics and MCTP, University of Michigan, Ann Arbor,
Michigan 48109 \\
$^{2}$ Institute for Quantum Optics and Quantum Information of the Austrian
Academy of Sciences, Austria\\
and Institute for Theoretical Physics, University of Innsbruck, A-6020
Innsbruck, Austria}
\date{\today }

\begin{abstract}
It is hard to stabilize the p-wave superfluid state of cold atomic gas in
free space due to inelastic collisional losses. We consider the p-wave
Feshbach resonance in an optical lattice, and show that it is possible to
have a stable p-wave superfluid state where the multi-atom collisional loss
is suppressed through the quantum Zeno effect. We derive the effective
Hamiltonian for this system, and calculate its phase diagram in a
one-dimensional optical lattice. The results show rich phase transitions
between the p-wave superfluid state and different types of insulator states
induced either by interaction or by dissipation.
\end{abstract}

\maketitle

The observation of the s-wave superfluid state in a Fermionic atomic gas
represents a remarkable breakthrough in the study of many-body physics with
ultracold atoms \cite{1}. The Feshbach resonance plays an important role in
those experiments, enhancing the interatomic interaction so that the
superfluid phase can be entered at a temperature that is experimentally
achievable. It is of great interest to realize the superfluid state with
other pairing symmetries as well. The p-wave superfluid state is the next
candidate for observation, and has attracted the interest of many
experimental and theoretical groups \cite{2,3}. The p-wave Feshbach
resonance has been recently observed in experiments \cite{2}, and can push
the single-component Fermi gas to the strongly interacting region and open a
door towards observation of the p-wave superfluid state in this system.
However, compared with the s-wave Feshbach resonance, a key difficulty with
the p-wave resonance is that the inelastic collision loss in this system is
typically large \cite{2,4,5}, which forbids thermalization of the gas within
the system lifetime \cite{5}.

In this paper, we dicuss how a dissipation-induced blockade mechanism can
stabilize the p-wave superfluid state in an optical lattice in the strongly
interacting region. Such a dissipation-induced blockade has been reported in
recent experiments to realize the Tonks gas \cite{6} or for simulation of
effective three-body interactions \cite{6a} with cold bosonic atoms. We
apply this mechanism to stabilize the single-component Fermion system in an
optical lattice in the presence of the p-wave Feshbach resonance. The p-wave
Feshbach resonance has been considered recently on single sites in a very
deep optical lattice \cite{4,7}. Here, instead, we focus on the many-body
physics by deriving an effective Hamiltonian for this system, taking into
account the atomic hopping in the reduced Hilbert space caused by the
dissipation-induced blockade. This effective Hamiltonian provides a starting
point to understand the quantum phases. We compute the phase diagram of the
system explicitly with well-controlled numerical methods for an anisotropic
lattice where the atom tunnelling is dominantly along one dimension. The
results show rich phase transitions between the p-wave superfluid state, a
dissipation-induced insulator state, the Mott insulator state, and different
kinds of metallic states. Although these results are obtained from
one-dimensional calculations, we expect these phases to correspond also to
similar phases in higher dimensions.

We consider a single-component Fermi gas near a $p$-wave Feshbach resonance.
If this strongly interacting gas is loaded into an optical lattice, many
different Bloch bands can be populated, in particular when the resonance is
broad (as it is the case for some recent experiments \cite{2}). However, we
can derive an effective single-band model for this system that is
independent of the interaction details. Following a strategy similar to the
s-wave Feshbach resonance case \cite{8}, we first analyze the local Hilbert
space structure on a single lattice site. When we have zero or one atom on
the site $i$, the states are simply denoted by $\left\vert 0\right\rangle
_{i}$ and $\left\vert a\right\rangle _{i}=a_{i}^{\dagger }\left\vert
0\right\rangle _{i}$, respectively. For the case of two atoms on a single
site, the exact two-body physics has recently been calculated \cite{7}, and
there are several two-body energy levels separated by an energy difference
of the order of the band gap. If we assume that the system temperature is
significantly below the band gap, only the lowest two-body state is
relevant. We refer to this state as a dressed molecule level, and denote it
by $\left\vert b\right\rangle _{i}=b_{i}^{\dagger }\left\vert 0\right\rangle
_{i} $. Note that the wave function of $\left\vert b\right\rangle _{i}$ in
general includes contributions from many of the original atomic orbitals
\cite{7}. It has a p-wave symmetry in space and is antisymmetric under
exchange of the two atoms in this dressed molecule.

If more that two atoms coming to a single site, different from the s-wave
case, the state will not be stable due to big three-particle inelastic
collision loss \cite{4}. At first sight, this seems to mean that the system
will become unstable. However, in a lattice, there is a subtle dissipation
induced blockade mechanism \cite{6} which forbids population of the unstable
three-particle state and thus stabilizes the whole system. The basic idea is
illustrated in Fig.~1. The three-particle state $\left\vert 3\right\rangle $
has a large bandwidth characterized by its inelastic collision rate $\gamma $
and an energy shift characterized by the on-site atom-dressed-molecule
interaction energy $\Delta _{3}$. If a single atom tunnel through a barrier
with a hopping rate $t$ to form this state $\left\vert 3\right\rangle $, the
probability of getting $\left\vert 3\right\rangle $ is given by $%
t^{2}/\left( \gamma ^{2}+\Delta _{3}^{2}\right) $ (similar to a
two-level transition with a detuning $\Delta _{3}\ $and a bandwidth
$\gamma $). So the net collisional loss of the system is bounded by
$\gamma _{\text{eff}}=\gamma t^{2}/\left( \gamma ^{2}+\Delta
_{3}^{2}\right) \leq t^{2}/\left( 2\Delta _{3}\right) $ no matter
how large the inelastic collision rate $\gamma $ is.
Near the Feshbach resonance, the atom-dressed-molecule interaction energy $%
\Delta _{3}$ is comparable with the lattice band gap (thus much larger than $%
t$) \cite{8a}, the net collisional loss $\gamma_{\text{eff}}$ is
therefore small
compared with the atomic hopping required to thermalize the system \cite%
{note}. The reduction of population in the noisy state $\left\vert
3\right\rangle $ is called the dissipation-induced blockade (or interpreted
as the quantum Zeno effect \cite{6}; the blockade is actually induced by
both dissipation and interaction when $\Delta _{3}$ and $\gamma $ are
comparable). Due to this mechanism one can achieve many-body thermal
equilibrium in a lattice even if there is big inelastic collision loss.

\begin{figure}[tbp]
\includegraphics[width=8cm]{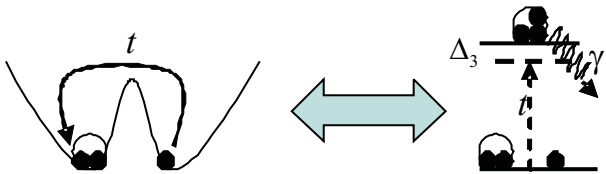}
\caption[Fig. 1 ]{(Color Online) Illustration of the
dissipation-induced blockade for multiple occupation (more than
two) of a single lattice site. Similar to a two level transition
with a detuning $\Delta _{3}$ and a decay rate $\protect\gamma$
for the target level, the effective hopping in this case is
suppressed by a factor  $t^{2}/(\protect\gamma^{2}+\Delta
_{3}^{2})$, where $t$ is the atomic tunneling rate.}
\end{figure}

Due to the dissipation-induced blockade discussed above, on each site we
have only three relevant levels: $\left\vert 0\right\rangle _{i},\left\vert
a\right\rangle _{i},\mbox{ and }\left\vert b\right\rangle _{i}$ as the low
energy configurations. The energy difference between the sates $\left\vert
a\right\rangle _{i}$ and $\left\vert b\right\rangle _{i}$ can be tuned with
the external magnetic field via the Feshbach resonance. We then take into
account the atomic hopping, which changes the level configurations of the
neighboring sites under the constraint that the atom number be conserved.
Thus, there are only three possible processes, as illustrated in Fig.~2.
Corresponding to these configuration changes, the general Hamiltonian for
this lattice system then takes the form

\begin{widetext}

\begin{eqnarray}
H &=&\sum_{i}\left[ (\Delta b_{i}^{\dagger }b_{i}-\mu
(a_{i}^{\dagger }a_{i}+2b_{i}^{\dagger }b_{i})\right]
 -\sum_{\left\langle i,j\right\rangle }P\left[
t_{1}a_{i}^{+}a_{j}+t_{2}(b_{i}^{+}-b_{j}^{+})a_{i}a_{j}+t_{3}b_{i}^{+}b_{j}a_{j}^{+}a_{i}+H.c.%
\right] P,
\end{eqnarray}%

\end{widetext}
where $\mu $ is the chemical potential and $\Delta $ is the energy detuning
of the dressed molecule controlled with the magnetic field. The value of $%
\Delta $ characterizes the on-site atomic interaction magnitude. The hopping
rates for the three processes illustrated in Fig.~2 are different, in
general, due to the contributions from different bands (because the dressed
molecule is a composite particle with population in multiple bands). The
hopping takes place in the low energy Hilbert space specified by the
projector $P\equiv \mathop{\displaystyle \bigotimes }\limits_{i}\left(
\left\vert 0\right\rangle _{i}\left\langle 0\right\vert +\left\vert
a\right\rangle _{i}\left\langle a\right\vert +\left\vert b\right\rangle
_{i}\left\langle b\right\vert \right) $, and the summation in Eq. (1) is
over all neighboring sites $\left\langle i,j\right\rangle $.

\begin{figure}[tbp]
\includegraphics[width=8cm]{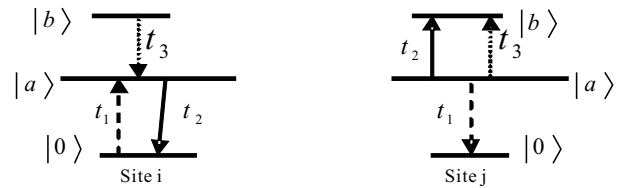}
\caption[Fig. 2]{(Color Online) Illustration of different atomic hopping
processes over the two neighboring sites described by the Hamiltonian (1),
where each site has only three possible level configurations.}
\end{figure}

The Hamiltonian (1), although much simplified compared with multi-band
models, is still complicated, and in general it does not allow exact
solutions. To understand some basic physical properties of the system, we
limit ourselves in the following to the one-dimensional lattice where the
atomic hopping along the other two dimensions are turned off with a high
lattice barrier. In this case, we can solve this model with well controlled
numerical simulations, and the results show rich phase transitions between
the p-wave superfluid state and different kinds of insulator and metallic
phases. We expect most of the phases found in the one-dimensional case have
counterparts in higher dimensions. In particular, the p-wave superfluid
state characterized by a quasi-long-range pairing order with diverging pair
susceptibility in one dimension can be easily stabilized to a true
long-range order if we allow weak tunneling between the one-dimensional
tubes \cite{9}.

In the numerical simulation, we use the iTEBD algorithm, a recently
developed method related to density matrix renormalization group techniques
\cite{10,11}, which allows direct calculation of the physical properties in
the thermodynamic limit. The algorithm has been shown to work with high
precision compared to known results for the Hubbard model \cite{12}. For
simplicity, we take the hopping rates $t_{1}=t_{2}=t_{3}=t$, and use $t$ as
the energy unit. Then we have effectively only two parameters, $\Delta$ and $%
\mu$ (in units of $t$) in the Hamiltonian (1). To figure out the complete
phase diagram with respect to these two parameters, we calculate $\partial
\left\langle H\right\rangle /\partial \Delta $ and $\partial \left\langle
H\right\rangle /\partial \mu $ as functions of $\Delta $ or $\mu $ for the
ground state of $H$, and use the characteristics of these curves to identify
the phase transition points.

\begin{figure}[tbp]
\includegraphics[width=8.0cm,height=4cm]{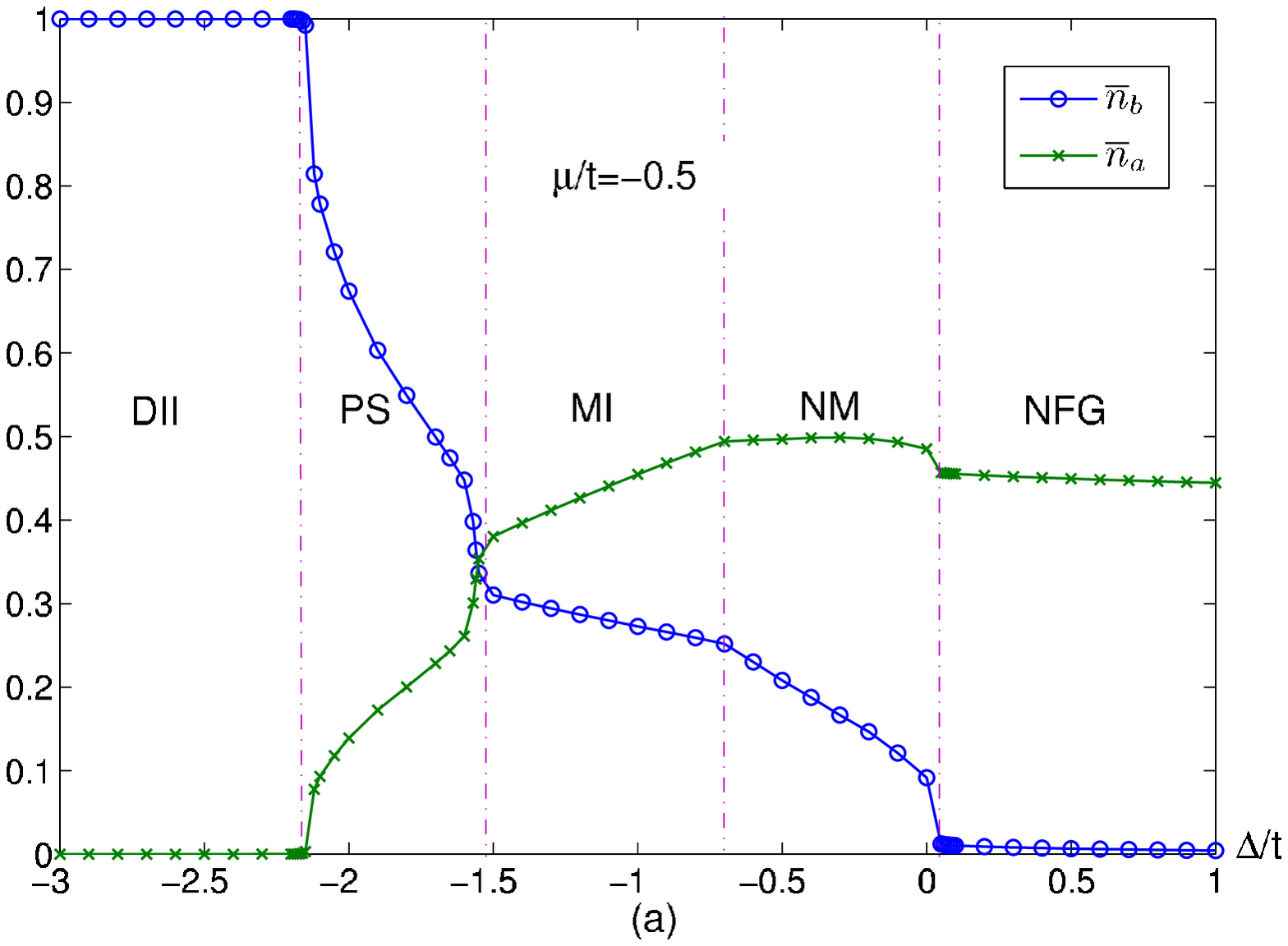} %
\includegraphics[width=8.0cm,height=4cm]{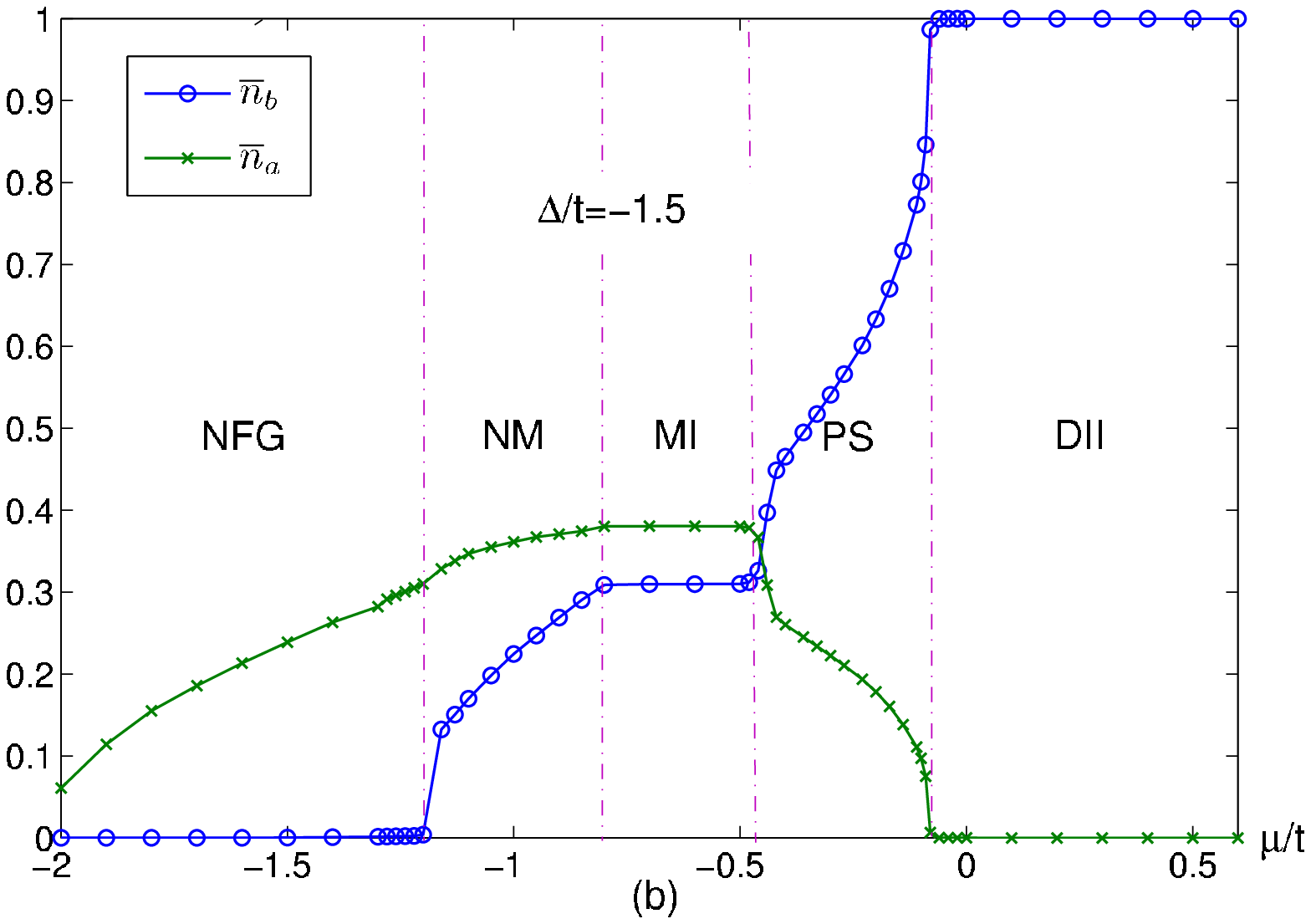}
\caption[Fig. 3]{(Color Online) The dressed molecule number $\overline{n}_{b}
$ (the double occupation probability) and the atom number $\overline{n}_{a}$
(the single occupation probability) shown as the function of $\Delta/t$
(with a fixed $\protect\mu/t=-0.5$) in Fig. (a) and $\protect\mu/t$ (with a
fixed $\Delta/t=-1.5$) in Fig. (b). The non-analyticities of these curves
signal a number of quantum phase transitions from the dissipation induced
insulator state (DII), to a p-wave superfluid state (PS), to a Mott
insulator state (MI), to a normal mixture state (NM), and finally to normal
Fermi gas phase (NFG). }
\end{figure}

In Fig.~3, we show $\overline{n}_{b}=\left\langle b_{i}^{\dagger
}b_{i}\right\rangle $ and $\overline{n}_{a}=\left\langle a_{i}^{\dagger
}a_{i}\right\rangle $ as functions of $\Delta $ and $\mu $. One can clearly
see several quantum phase transitions from this figure. First, with a fixed
chemical potential $\mu =-0.5t$, one has $\overline{n}_{b}=1$ and $\overline{%
n}_{a}=0$ with a large negative detuning $\Delta $ (corresponding to strong
attractive atomic interaction). In this case, each site is doubly occupied
with two atoms. More than two atoms cannot move to the same site because of
the dissipation-induced blockade. So this is an insulator phase stabilized
by the dissipation. As one increases $\Delta $ with $\Delta >-2.15t$, the
number of atoms on each site begins to fluctuate. If one looks at the pair
correlation $\left\langle b_{i}^{\dagger }b_{j}\right\rangle $, it shows
quasi-long-range behavior with slow algebraic decay. In Fig.~4(a), we show
this correlation in the $k$-space, defined as $P_{k}=\left( 1/N\right) %
\mathop{\displaystyle \sum }\limits_{\rho =0}^{N}\left\langle b_{i}^{\dagger
}b_{i+\rho }\right\rangle e^{ik\rho }$. The correlation $P_{k}$ is peaked
sharply at $k=0$. This corresponds to the $p$-wave superfluid phase. The $p$%
-wave character is inherited from the $p$-wave symmetry of the dressed
molecule in space $b_{i}$ (or the atomic pair on the same site). The $p$%
-wave nature of the pairing is also manifested in the atomic pair
wavefunction at different sites $\left\langle a_{i}a_{j}\right\rangle $,
which is obviously antisymmetric under exchange of the sites. In the
one-dimensional case, the $p$-wave superfluid state is characterized by a
quasi-condensate of the atomic pairs with a diverging pairing
susceptibility. If one allows weak coupling between the one-dimensional
tubes in the optical lattice, the $p$-wave quasi-condensate can easily be
stabilized into a real condensate with a true long range pairing order.

\begin{figure}[tbp]
\includegraphics[width=4.25cm]{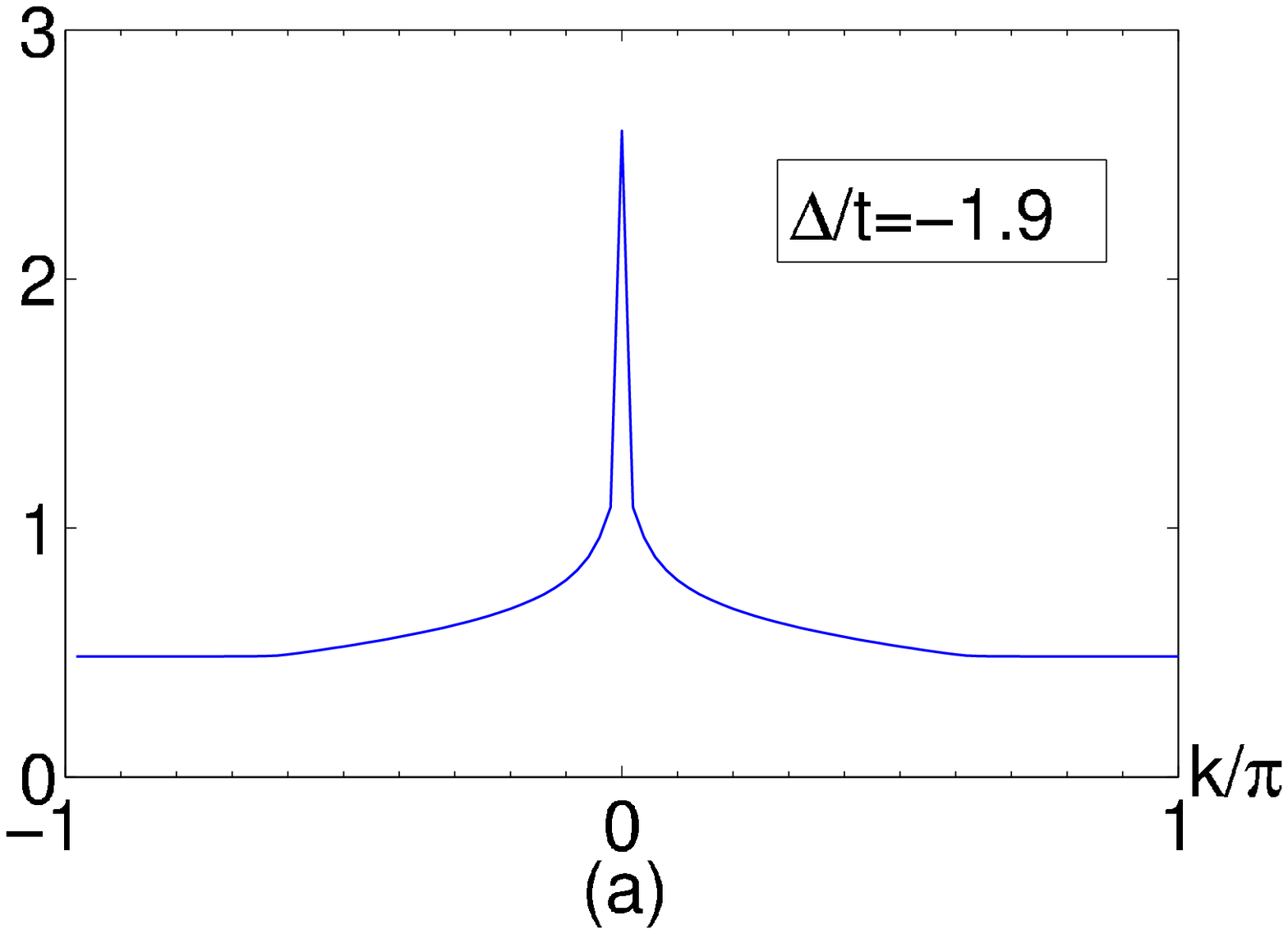} %
\includegraphics[width=4.25cm]{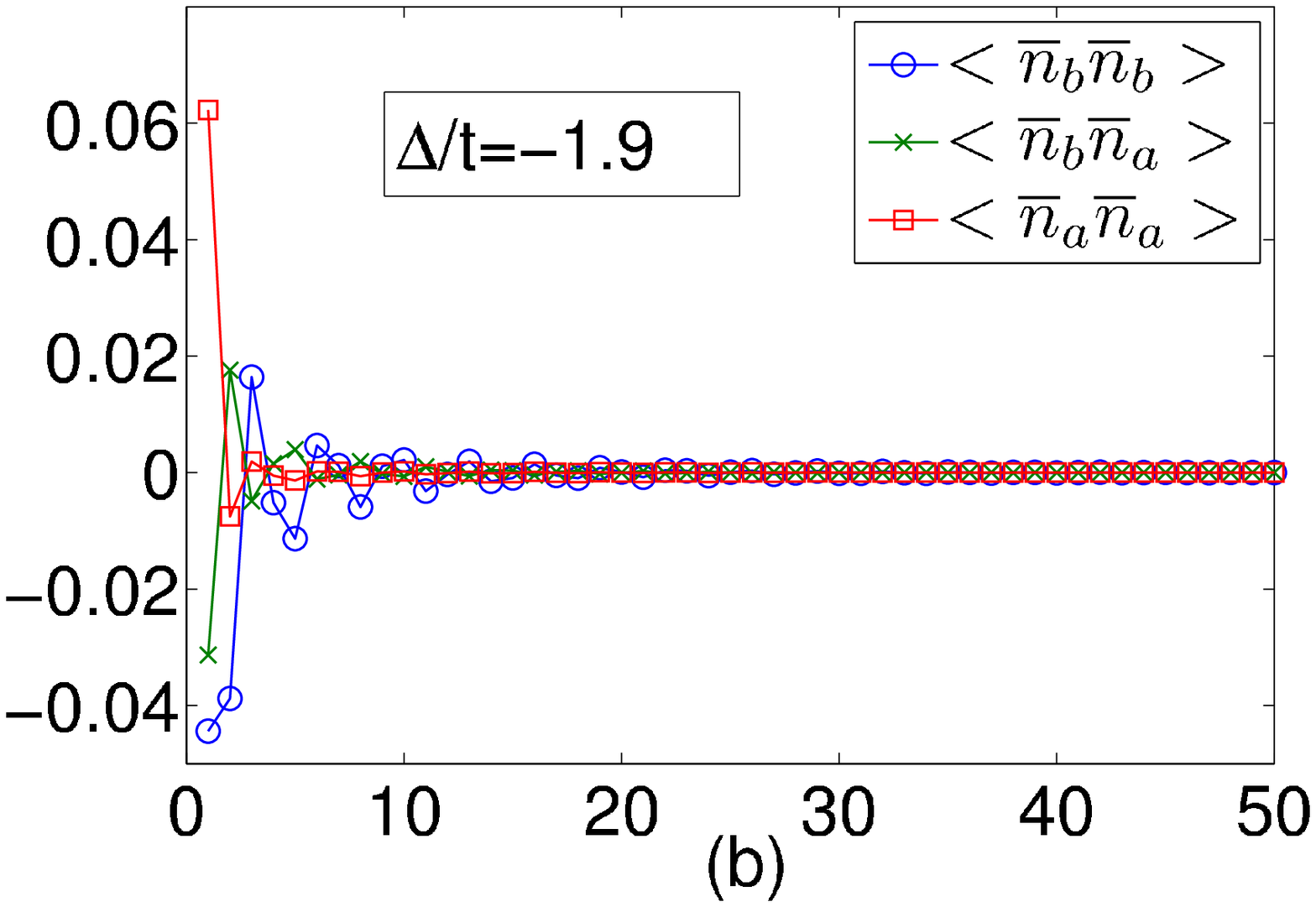} %
\includegraphics[width=4.25cm]{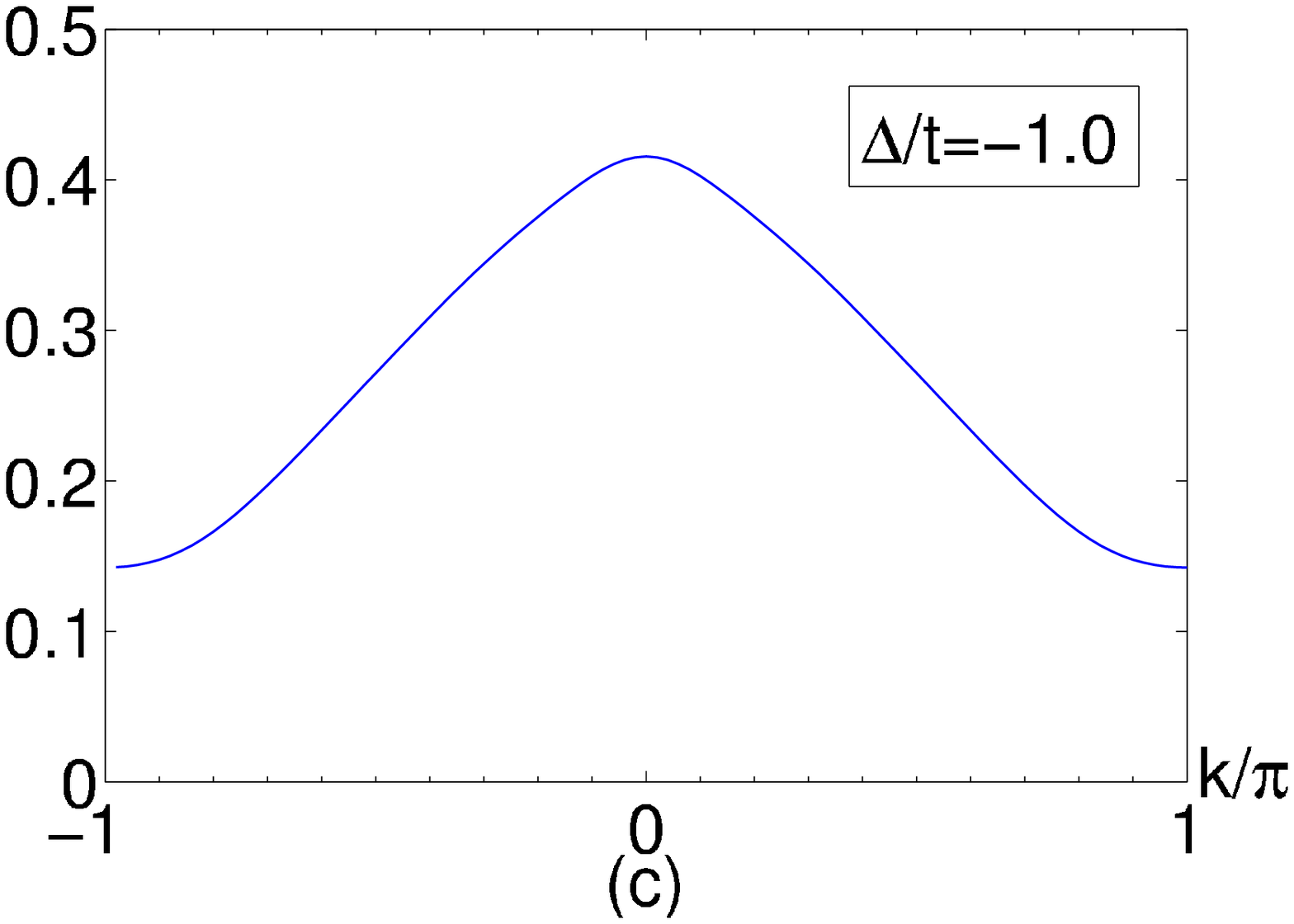} %
\includegraphics[width=4.25cm]{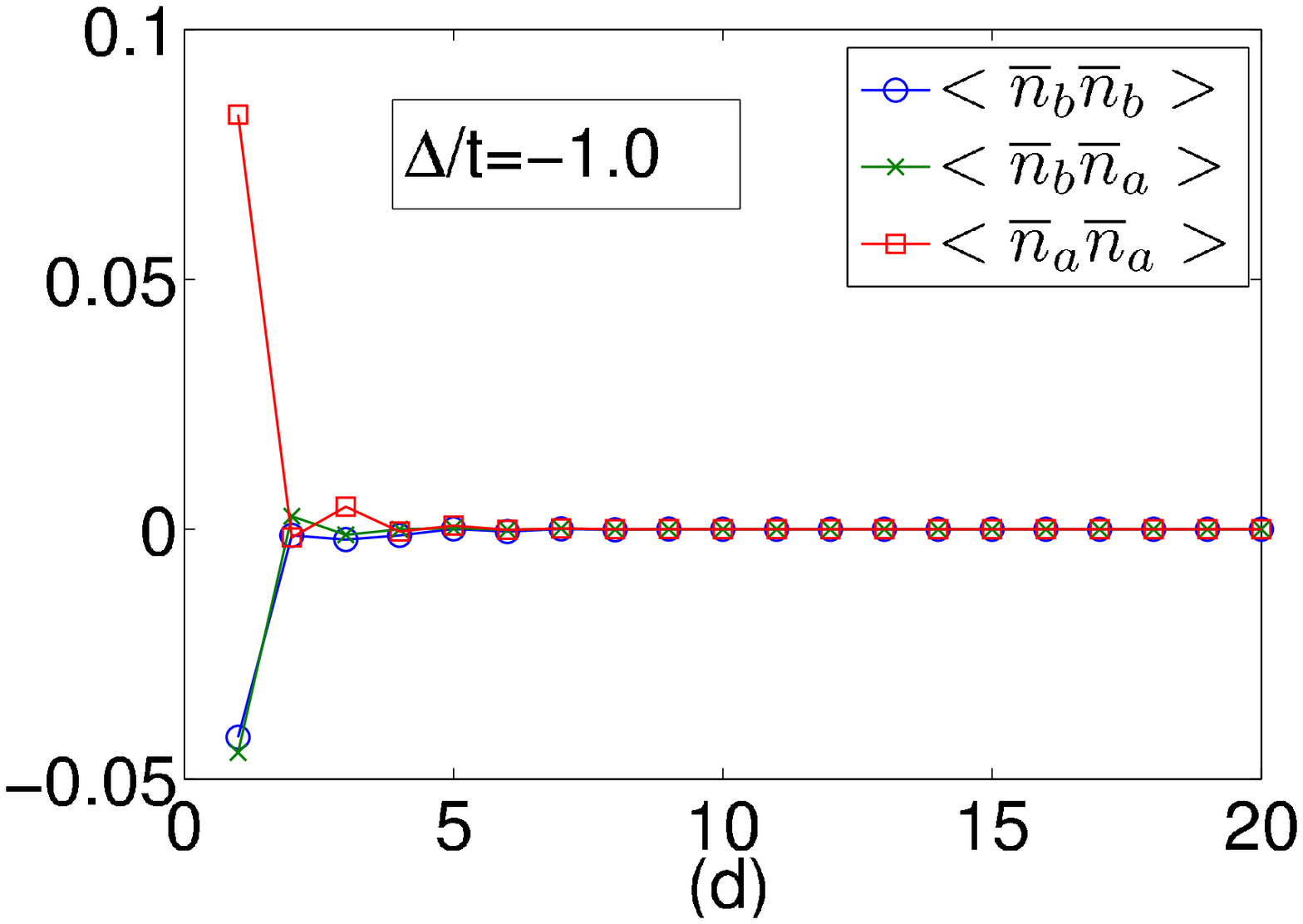} %
\includegraphics[width=4.25cm]{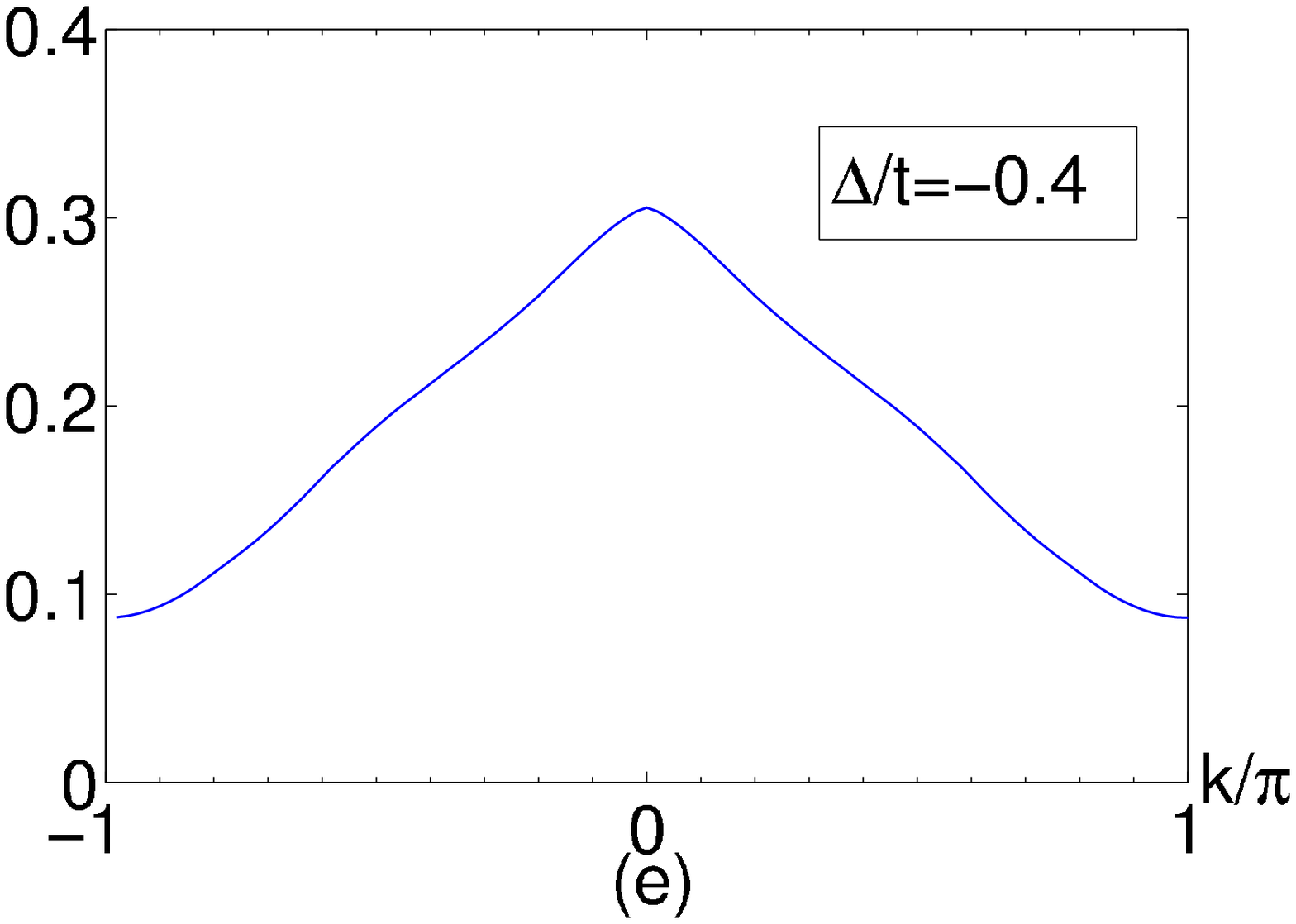} %
\includegraphics[width=4.25cm]{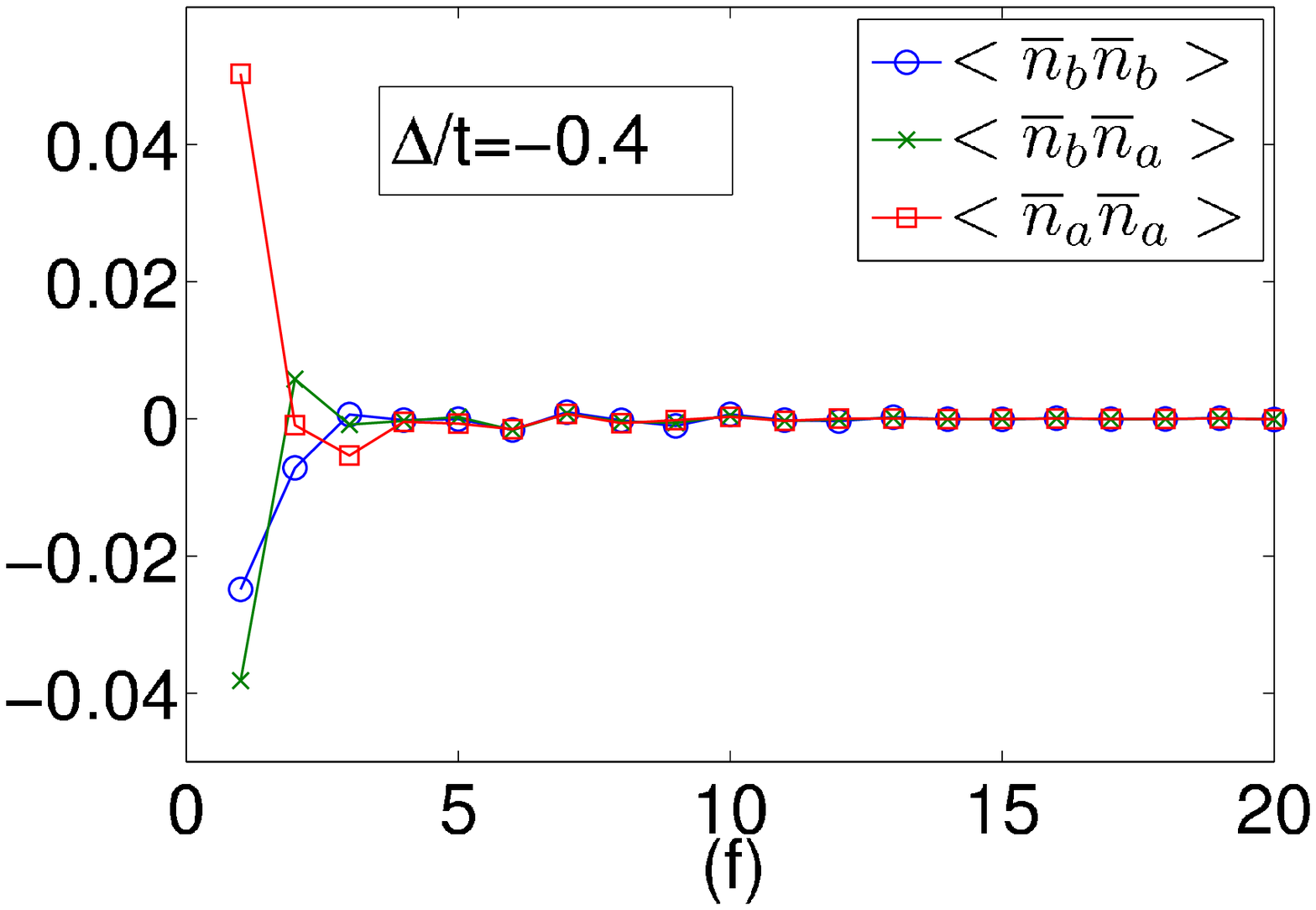}
\caption[Fig. 4]{(Color Online) The correlation functions shown for the
p-wave superfluid phase ((a) and (b)), the Mott insulator phase ((c) and
(d)), and the normal mixture phase ((e) and (f)). Figures (a), (c) and (e)
show in $k$-space the Fourier transform $P_{k}$ of the pair correlation $%
\left\langle b_{i}^{\dagger }b_{j}\right\rangle $. Figures (b), (d) and (f)
show the charge density correlations of $\left\langle n_{bi}
n_{bj}\right\rangle $, $\left\langle n_{bi} n_{aj}\right\rangle $ and $%
\left\langle n_{ai}n_{aj}\right\rangle $ in real space. We take $\protect\mu%
/t=-0.5$ for all the figures. }
\end{figure}

If one further increases $\Delta $ in Fig.~3(a), one enters a phase where
the total particle number per site $n=2\overline{n}_{b}+\overline{n}_{a}$ is
fixed at $1$ (although the double occupation probability $\left\langle
b_{i}^{\dagger }b_{i}\right\rangle $ varies with $\Delta $). This is a Mott
insulator state with a finite gap to charge excitations. This can be seen
more clearly in Fig.~3(b), where we fix the detuning $\Delta $, and show $%
\overline{n}_{b}$ and $\overline{n}_{a}$ as functions of the chemical
potential $\mu $. For this phase, the number density does not vary with $\mu
$, so the system is incompressible as one expects for a Mott insulator
phase. The correlation functions for this phase is shown in Fig.~4, where
both the pair correlation and the charge density wave correlations are of
short range. As one further moves to the right side in Fig.~3(a), there are
two other phases: the normal mixture (NM) and the normal free gas (NFG).
Both of these two phases are of a metallic nature with a finite
compressibility (see Fig.~3(b)). The difference is that in the normal
mixture phase, some sites are doubly occupied (with a finite $\left\langle
b_{i}^{\dagger }b_{i}\right\rangle $). Several kinds of correlation
functions for the normal mixture phase are shown in Fig. 4, and all of them
decay rapidly with distance. In the NFG phase, the double occupation
probability $\left\langle b_{i}^{\dagger }b_{i}\right\rangle $ reduces to
zero, and one has a conventional single component free Fermion gas.

\begin{figure}[tbp]
\includegraphics[width=8.5cm,height=5cm]{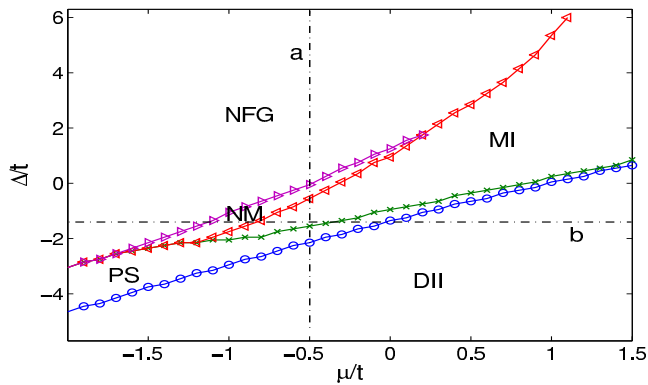}
\caption[Fig. 5]{(Color Online) The complete phase diagram of the system
versus two parameters $\protect\mu/t$ and $\Delta/t$. The five different
phases are marked with the same notation as in Fig. 3. The black dashed
lines a and b correspond to the parameters taken in Fig. 3(a) and Fig. 3(b).
}
\end{figure}

We have calculated the phase transition points for all $\Delta $ and $\mu $,
and the result is summarized in Fig.~5 to give a complete two-parameter
phase diagram. The five different phases are marked there. Under real
experimental conditions, there is typically a weak global harmonic trap in
addition to the optical lattice potential. As one moves from the trap center
to the edge, the effective chemical potential $\mu $ gradually decreases
under the local density approximation. So with a fixed interaction parameter
$\Delta $, a line in the phase diagram of Fig.~5 across different $\mu $
gives the phase separation pattern of the Fermi gas in a harmonic trap. With
a large positive $\Delta $, one has a Mott insulator state in the middle,
surrounded by a normal Fermi gas. The phase transitions are most rich for
small $\left\vert \Delta \right\vert $, where one can cross all the five
different phases from the trap center to the edge. For large negative $%
\Delta $, the region of the $p$-wave superfluid state increases, but the
Mott insulator and the normal mixture states eventually disappear when $%
\Delta <-3t$. As the p-wave superfluid state has a large stability region in
the phase diagram, such a phase can be prepared experimentally by
adiabatically ramping the Hamiltonian parameters following certain
trajectories that suppress the three-particle occupation \cite{6a}.

In summary, we suggest that the p-wave superfluid state near a Feshbach
resonance can be stabilized in an optical lattice through a
dissipation-induced blockade mechanism. We have derived the effective
Hamiltonian to describe this strongly interacting system, taking into
account the restriction of the Hilbert space due to this blockade mechanism.
We solve the Hamiltonian in the anisotropic one-dimensional lattice through
exact numerical calculations, and the result suggests rich phase transitions
between the p-wave superfluid state and several kinds of insulator or
metallic phases.

We thank Ignacio Cirac, Jason Kestner, Wei Zhang, and Mikhail Baranov for
helpful discussions. This work was supported by the AFOSR through MURI, the
DARPA, the IARPA, and the Austrian Science Foundation through SFB FOQUS.


\begin{thebibliography}{99}
\bibitem{note} As an example, for $^{40}$K, the measured $1/e$ life-time is $%
\sim 1$ms \cite{2}. If we scale this value to the lattice case with
a typical depth $V_{0}\sim 10E_{r}$, the decay rate $\gamma $ can be
a hundred times larger than the tunneling rate $t$, which leads to
small effective collisional loss $\gamma _{\text{eff}}$ even without
the detuning $\Delta _{3}$ caused by interaction.
\end{thebibliography}
\end{document}